# Next steps of quantum transport in Majorana nanowire devices

Hao Zhang[1,2,3*], Dong E. Liu[1,3], Michael Wimmer[4] and Leo P. Kouwenhoven[4,5]

[1]*State Key Laboratory of Low Dimensional Quantum Physics,*

*Department of Physics, Tsinghua University, Beijing, 100084, China*

[2]*Beijing Academy of Quantum Information Sciences, Beijing 100193, China*

[3]*Frontier Science Center for Quantum Information, Beijing, 100084, China*

[4]*Qutech and Kavli Institute of Nanoscience, Delft University of Technology, 2600 GA Delft, The Netherlands*

[5]*Microsoft Station Q Delft, 2600 GA Delft, The Netherlands*

**Majorana zero modes are localized quasiparticles that obey non-Abelian exchange statistics. Braiding Majorana zero modes forms the basis of topologically protected quantum operations which could in principle significantly reduce qubit decoherence and gate control errors in the device level. Therefore, searching for Majorana zero modes in various solid state systems is a major topic in condensed matter physics and quantum computer science. Since the first experimental signature observed in hybrid superconductor-semiconductor nanowire devices, this field has witnessed a dramatic expansion in material science, transport experiments and theory. While making the first topological qubit based on these Majorana nanowires is currently an on-going effort, several related important transport experiments are still being pursued in the near term. These will not only serve as intermediate steps but also show Majorana physics in a more fundamental aspect. In this perspective, we summarize these key Majorana experiments and the potential challenges.**

A strong spin-orbit coupled semiconductor nanowire (NW) contacted by a superconductor (S) can be turned into a topological superconductor by applying a magnetic field along the wire[1-3]. Majorana zero modes (MZM) are predicted to form at the topological phase boundary (*i.e.* the wire ends), and can be detected in the tunneling conductance as a zero bias peak (ZBP)[4-6]. Following this theoretical proposal, a ZBP at finite magnetic field was first observed in an InSb nanowire covered by a NbTiN superconductor in 2012 [7]. While the gate and magnetic field dependence of this ZBP is consistent with Majorana theory and similar ZBPs were quickly reproduced by other groups[8-11], many alternative explanations with trivial origins were proposed soon after[12-16]. Moreover, the induced superconducting gap in these original experiments showed finite sub-gap conductance in the low conductance tunneling limit. This soft gap can spoil Majorana signatures and more importantly destroy the topological protection. Further study showed that the soft gap and most of the alternatives are related to disorders at the superconductor-nanowire interface[17]. Engineering high quality clean interface with better control on nanowire device fabrication lead to quantized conductance plateaus[18-20], a hallmark for ballistic one dimensional system. Clean and more robust ZBPs in these ballistic nanowire devices provide confidence in ruling out most of the alternative explanations that invoke disorder[21]. More importantly, epitaxial grown of superconductor (Al) directly on InAs and InSb nanowires leaves an atomic flat interface[22, 23]. This material breakthrough resulted in a hard superconducting gap even at finite magnetic field together with improved ZBP quality[24, 25], solving the soft gap problem. Finally, the ZBP height was found to be quantized at $2e^2/h$ [26, 27], closing one chapter in tunneling spectroscopy based on the simplest normal lead (N)-NW-S device[28, 29].

---

[*] email: hzquantum@mail.tsinghua.edu.cn

These series of experimental and material breakthroughs since 2012, together with deep theoretical understanding, make Majorana nanowires one of the most promising platforms to realize non-Abelian statistics and topological quantum computing[30, 31] through a braiding experiment[32-34]. While much efforts have been invested along this roadmap, other Majorana transport experimental schemes, which could not only serve as intermediate steps to realize non-Abelian braiding statistics but also reveal its exotic fundamental physics, still remain as an important quest. Here we summarize several key schemes with the basic Majorana signature and their potential challenges. These schemes, not experimentally fully achieved yet, can establish more comprehensive aspects of Majorana physics and guide the braiding experiment. We note that this Perspective is by no means to cover all proposed schemes in literature, but only those with simple device designs (relatively easy to achieve in the near-future). Some of these schemes are currently being pursued in labs with even some preliminary results.

**Peak-to-dip transition in quantized Majorana conductance.** Fig. 1a shows the schematic setup for a first experiment that extends the original ZBP measurements: a typical N-NW-S device with two gates tuning the electro-chemical potential and tunnel barrier, respectively. When MZMs form at the application of a magnetic field, they give rise to quantized ZBPs (black curves in Fig. 1b) when the tunnel barrier only allows one spin-polarized channel to transmit, corresponding to a normal state conductance (*i.e.* when the bias voltage is tuned outside the superconducting gap) lower than $e^2/h$. Lowering the tunnel barrier will eventually occupy the second spin-polarized channel. In this case, conductance through one channel will stay blocked; in the typical case where the second channel has spin opposite to the first one (when Zeeman energy is smaller than the transverse confinement energy), this can be understood as a spin selection rule of MZMs[4, 35, 36]. In experiment, the zero bias conductance will thus remain quantized at $2e^2/h$ with increasing tunnel barrier transmission (normal state conductance), resulting in a quantized zero bias dip (red curves in Fig. 1b). Lowering the barrier further adds a third channel as a background, which can eventually push the zero bias conductance above $2e^2/h$.

Recently, theory showed that quantized ZBP can be mimicked by partially separated Andreev bound states (ps-ABS)[37] or quasi-Majorana states[38, 39] (interestingly, these states, though trivial, mimic MZMs to a degree that even a braiding experiment may be possible[38]). The key idea behind is to create two Majorana states with opposite spins near the tunnel barrier with spatial overlap. Due to smooth potential inhomogeneity[40, 41] which prevents the large-momentum scatterings, these two Majorana states from two separated Fermi surfaces have negligible coupling. If the tunnel barrier only has one spin polarized channel occupied, the normal lead can only couple to one Majorana state. As a result, the device will show quantized ZBPs. However, lowering the barrier to have the second channel with opposite spin occupied can couple the normal lead with the second Majorana state as well. In this case, the conductance contributed by two quasi-Majoranas will be a $4e^2/h$ ZBP instead of $2e^2/h$ zero bias dip. Therefore, the peak-to-dip transition in quantized Majorana conductance can reveal its spin selection property and be a unique signature to rule out the quasi-Majorana explanation[35, 38]. Experimentally, the quantized ZBP has been observed[26]. The next-step is to observe the transition from the peak to the quantized zero bias dip by lowering the barrier further. This can be achieved by increasing the capacitive coupling of the tunnel-gate, e.g. replacing the side-gates with a wrap-around-gate.

Besides the peak to dip transition, another experiment could be performed in the simplest N-NW-S device by making the N electrode highly resistive (comparable to $h/e^2$) while the contact is still Ohmic. By introducing this strong Ohmic dissipation in the probe electrode, Majorana ZBP shows non-trivial temperature scaling, while non-Majorana ZBP splits into two peaks as reducing temperature [42].

**Non-local Majorana gate effect.** The second experiment is to add an additional gate to tune the electro-chemical potential as shown in Fig. 2a, c. The gate close to the tunnel barrier is called local gate while the remote one is called non-local gate since the tunneling spectroscopy mainly detects the local density of states (LDOS) near the tunnel barrier. Tuning the non-local gate can move the remote MZM close to the tunnel barrier, hybridize with the local MZM and split the ZBP (Fig. 2b, d)[43]. This device thus detects the non-local property of MZMs[44, 45] by adding a non-local element (gate). Non-local gate dependence of ZBPs can, to a large extent, rule out ps-ABS and quasi-Majorana states which are localized near the tunnel barrier and thus not tunable by non-local gate. A careful control experiment is needed to rule out the cross capacitance coupling between the non-local gate and the local nanowire region. This can be achieved by increasing the local gate length and verifying that non-local gate has no effect on trivial states[46, 47] localized near the local nanowire region. Combined with systematic self-consistent electrostatic simulations[48-50], this 'non-local gating' experiment could reveal how the electro-gate moves Majorana states in real space and further help on extracting important system parameters like the Majorana wave function size. Besides the non-local gate, another way to tune the coupling between two Majorana states is through magnetic flux by contacting the nanowire with a superconducting loop[51].

**Correlation and three terminal Majorana device.** To fully reveal the non-local property of MZM, a true non-local measurement can be conducted in a three-terminal device (N-S-N). Fig. 3a shows the setup for such a non-local correlation experiment. Measuring the $dI_1/dV$ and $dI_2/dV$ from the nanowire's two ends can detect the two LDOS simultaneously. MZMs always showing up in pairs guarantees that the appearance of two ZBPs from the two ends and their splitting (Majorana oscillations) should be correlated in all parameter space, *i.e.* gates and magnetic field. The ZBP heights and widths depend on the local tunnel barrier and does not need to be correlated. One important requirement here is that the superconducting part of the wire needs to be sufficiently long [52] (much longer than the spatial distribution of a trivial Andreev bound state). Otherwise $dI_1/dV$ and $dI_2/dV$ may end up in detecting the same trivial state from two sides, mimicking a correlation signature. Correlation experiment can exclude the trivial Andreev bound state (ABS) explanation to a large extent. However, fine tuning the two tunnel barriers may also lead to ABS induced ZBPs showing up at the same magnetic field. To rule out this case, the robustness of ZBP correlations needs to be tested by varying magnetic field and voltages on all different gates. In addition, if one wire end has a quantum dot or smooth potential inhomogeneity, the two ZBPs from $dI_1/dV$ and $dI_2/dV$ may not show up at the same magnetic field due to the interruption of localized ABS [53]. A true correlation experiment on a long Majorana nanowire can demonstrate that Majorana zero modes are indeed correlated pairs at the two ends of a topological superconductor.

Another experiment can be conducted on the same device with a slightly different measurement circuit (Fig. 3c) for the detection of non-local crossed Andreev reflection processes[54]. While the local conductance $dI_{local}/dV$ (Fig. 3b) only probes the nanowire region near the tunnel barrier, the non-local conductance $dI_{non-local}/dV$ reveals the induced gap information of the entire proximitized nanowire if the wire is longer than the superconducting phase coherence length. In this long-wire regime, any local feature below the induced gap is fully suppressed in the non-local transport. More importantly, the non-local conductance is an odd function of bias voltage (with some parts having negative differential conductance) near the gap closing-reopening topological phase transition point (dashed line in Fig. 3d). This current rectifying effect is due to crossed Andreev reflection and can serve as a more reliable measure of topological phase transition, since localized trivial Andreev bound states due to potential inhomogeneity[46] can disturb the gap closing point in the local conductance, but not in the non-local conductance. Thus combining the correlation measurement with the crossed Andreev measurement in a *long-wire* device would allow to correlate the appearance of the Majorana ZBP with a gap closing.

**Majorana T-shape device for local density of states.** Although the major Majorana experiment activities focus on LDOS at the Majorana wire ends, a T-shape structure can be used to detect the wire bulk. Fig. 4a shows such a device where the side electrode can probe the LDOS in the wire bulk through the extra nanowire 'leg'. Since the MZMs are localized at the ends, the side probe is only able to probe gap closing and reopening with no ZBPs when sweeping magnetic field[55]. In the meantime, ZBP can be detected with end electrodes where the gap closing-reopening feature may not be visible due to small coupling to these bulk states. Recently, direct deposition of multiple probes on a single wire has been used to perform similar measurement where the superconducting gap is relatively soft and ZBP-height is small [56]. Moreover, the tunnel barrier is not gate tunable, making the device not fully functional: tunnel-gate dependence is an important experimental knob on the data interpretation and differentiation of trivial Andreev states. The biggest concern is probably the almost unavoidable potential inhomogeneity introduced by these side probes in the nanowire bulk[57-59]. This potential fluctuation can 1) create trivial Andreev bound states[39, 46, 60-62]; 2) easily break the topological superconducting region due to the small topological gap size. Both effects are detrimental for detecting MZMs. One possible solution is to apply negative enough gate voltage to push the electron wave-function close to the superconducting film[47-50, 63], and thus far away from the potential inhomogeneity located mostly near the T-junction[64]. This experiment can provide the spatial information of the LDOS in a topological superconductor wire.

**Majorana-Fu teleportation.** Tunneling spectroscopy can mainly reveal the local wave function information of MZMs. To study their exotic non-local feature (phase), a Majorana interferometer structure, initially proposed by Fu[65], can be implemented as shown in Fig. 5a [34]. A piece of superconducting island on the nanowire with finite charging energy forms a Majorana island[66]. A magnetic field can drive the Coulomb blockade of the island from 2$e$-periodic oscillations (Cooper pair) to 1$e$-periodic oscillations (coherent single electron 'teleportation', Fig. 5c), where MZM brings the odd parabola down to zero in energy diagram as shown in Fig. 5b. Now adding a coherent nanowire channel as a reference, electrons from the source have two coherent paths to reach the drain: the reference channel and the Majorana island through non-local teleportation. The interference of these two paths can be measured by Ahronov-Bohm (AB) effect with a magnetic flux through the loop. The key experimental signature is a π phase shift of the AB interference when changing the Majorana island parity as shown in Fig. 5d. Recent experiment [67] has realized such a device structure with π phase shift observed for the coherent single-electron transport, providing a good start. However, π phase shift is also generously observed in quantum dot based AB loops[68] and superconducting interference devices[69] when switching the dot/island parity. The field rotation dependence in Ref [67] also suggests additional non-Majorana mechanisms for those short islands. To differentiate from the trivial case, the Majorana island needs to be sufficiently long to suppress the trivial incoherent tunneling processes mediated by Andreev bound states[67, 70-73]; on the other hand, Majorana non-local teleportation still remains phase coherent. Regarding the wire length in reality, the requirement of finite charging energy also sets an upper bound on the island length. A π phase shift of AB oscillation in a *long-island*-based Fu-interferometer could reveal the coherence and parity change of two spatially well separated MZMs.

**Topological Kondo effect.** The final experiment we would like to discuss is the topological Kondo effect, initially proposed by Beri and Cooper[74]. This effect requires $N \geq 4$ MZMs on a superconducting island with finite charging energy, among which $M \geq 3$ MZMs are tunnel coupled to normal leads. The simplest setup is shown in Fig. 6a. Two parallel nanowires, each holding two MZMs, are contacted by an s-wave superconductor with three MZMs tunnel coupled to normal leads. The linear conductance, $G_{12} = \frac{dI}{dV}\big|_{V \to 0}$, is obtained by applying a small voltage excitation $V$ on lead-1 and measuring the current $I$ on lead-2 with the third lead grounded. The conductance shows Coulomb blockade oscillations by varying the gate voltage due to charging energy. Tuning the gate into a Coulomb blockade valley, the low-energy physics can be captured by an effective Kondo-type Hamiltonian with $SO(M)$ symmetry [75-77]. For a general case with spatially well separated MZMs (Fig. 6b), Fig. 6c shows the unique temperature dependence of the conductance with a logarithmic behavior and a non-trivial power-law behavior. As the temperature decreases, the conductance shows a crossover from a weak coupling trivial fixed point at high temperature to a strong coupling $SO_2(M)$ non-Fermi-liquid (NFL) fixed point at low temperature, where the linear conductance saturates at $G_{12} = 2e^2/Mh$. This crossover energy scale is the Kondo temperature $T_K$, which depends on the system parameters (e.g. coupling strength, charging energy etc). In Fig. 6d, a finite hybridization (overlap) $h_{34}$ between MZMs $\gamma_3$ and $\gamma_4$ is introduced[78, 79]. As a result in Fig. 6e, the conductance will initially try to reach $G_{12} = 2e^2/Mh$ when decreasing temperature, similar to Fig. 6c, as if the temperature is high enough to not 'feel' this hybridization energy $T_h = T_K(h_{34}/T_K)^{M/2}$. Below $T_h$, the hybridization between MZMs $\gamma_3$ and $\gamma_4$ becomes relevant and effectively removes these from the low-energy physics. The system's MZM number is thus decreased by 2, and the conductance saturates to a different value $G_{12} = 2e^2/(M-2)h$. This experiment will demonstrate that four Majorana zero modes can non-locally form a topological degeneracy, which is the basis of a topological qubit. Therefore, the topological Kondo device not only can provide a clear test of non-local Majorana quantum dynamics, but also shares the same device structure of a Majorana qubit design for future studies[32].

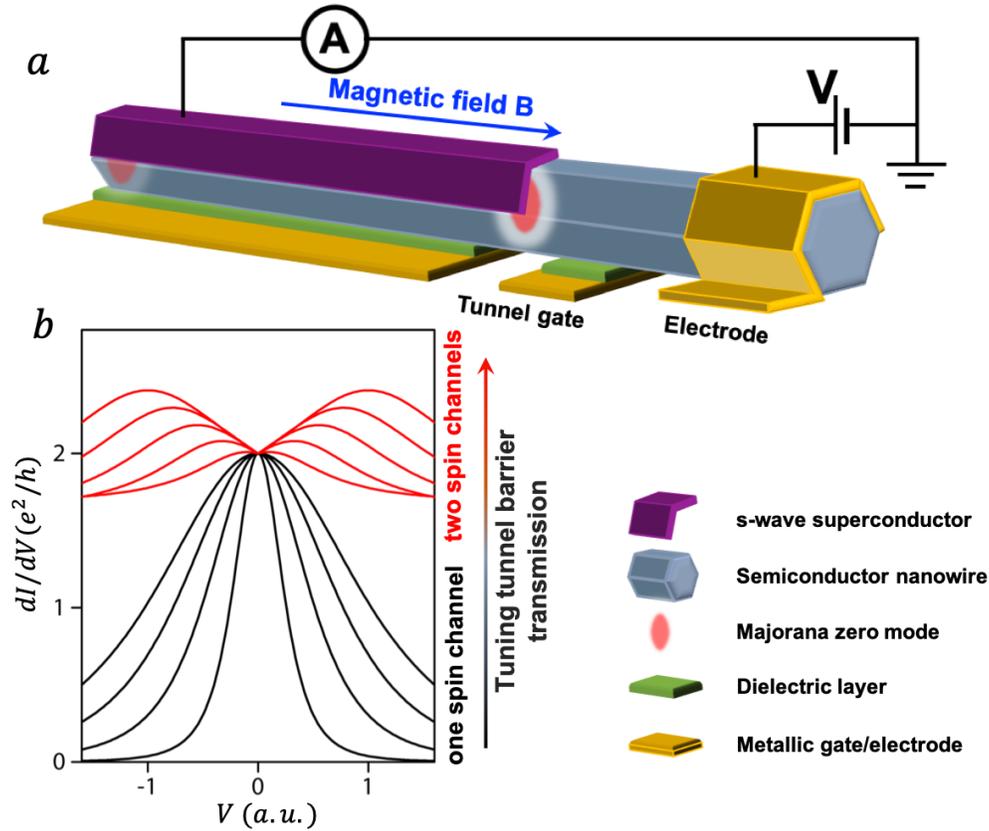

**Figure 1 | Majorana device for tunnel gate control**. **a**, Schematic experimental setup of the simplest Majorana nanowire device for tunneling spectroscopy measurement. **b**, Schematic plot of differential conductance $dI/dV$ as a function of bias voltage $V$ for different tunnel barrier transmission controlled by the tunnel gate. Increasing the tunnel barrier transmission, the d$I$/d$V$ shows a peak (black) to dip (red) transition with the zero bias conductance always quantized at $2e^2/h$, at sufficiently low temperature. The zero bias peaks (black) correspond to the case with only one spin-resolved channel in the tunnel barrier. The Majorana conductance still remains quantized for the two spin-resolved channels (red) due to the spin selection process of MZM. All the plots and devices in this paper are only for schematic purpose and not to scale.

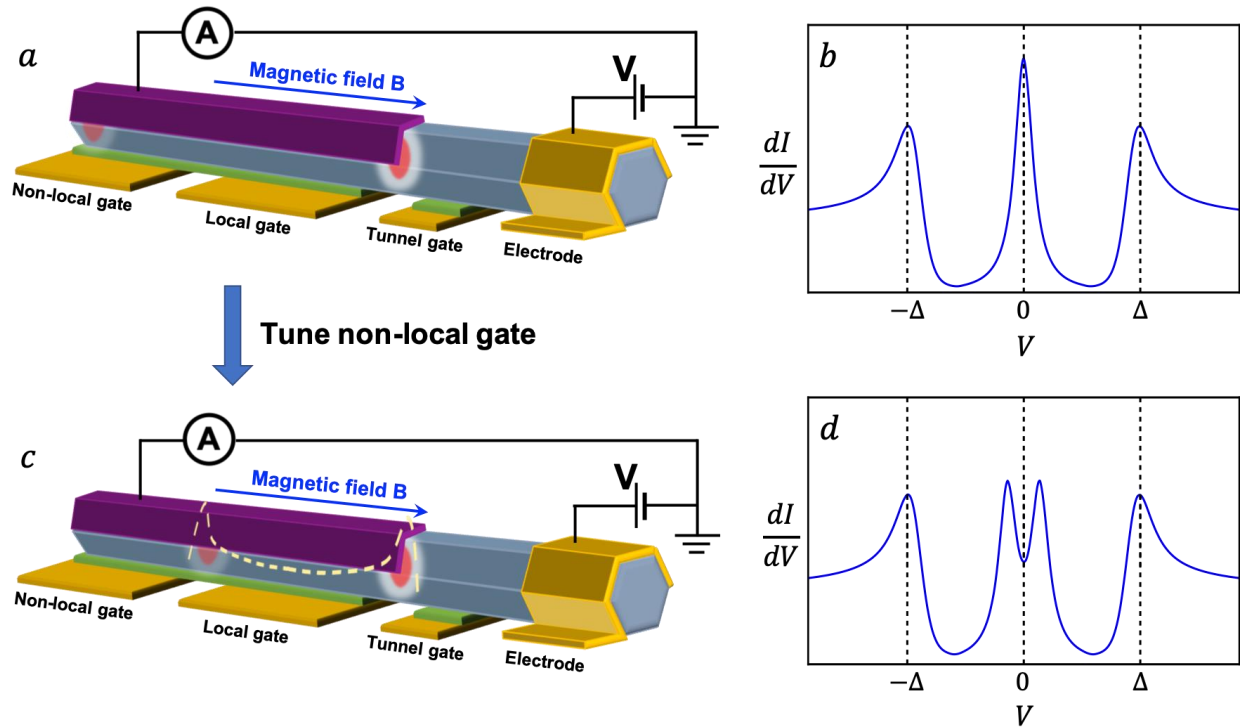

**Figure 2 | Majorana device for non-local gate control**. **a**, **c**, Schematic experimental setup with two separated gates tuning the electro-chemical potential. By fixing the local-gate and tuning the non-local gate, the left (remote) MZM moves to the right and hybridizes with the right (local) MZM. Therefore, the d$I$/d$V$, which detects the local density of state (LDOS) near the tunnel barrier, shows a peak (**b**) to split-peak (**d**) transition. $\Delta$ indicates the superconducting gap.

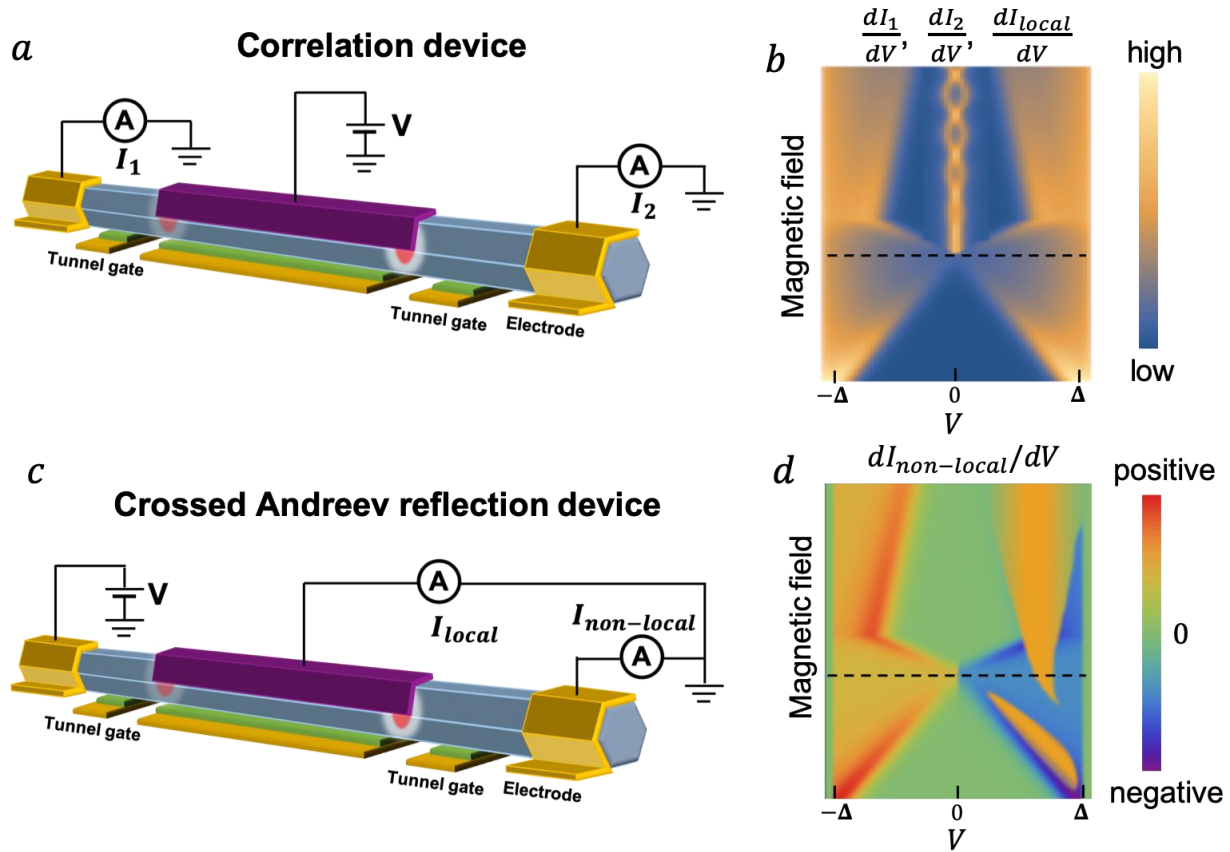

**Figure 3 | Majorana correlation and crossed Andreev reflection devices. a,** Schematic experimental setup of a Majorana correlation device to measure the currents $I_1$ and $I_2$ from the two ends, separately. **c,** Crossed Andreev reflection device to measure the local and non-local currents $I_{local}$ and $I_{non-local}$. **b,** Schematic plot of $dI_1/dV$, $dI_2/dV$, and $dI_{local}/dV$ indicate the correlated ZBP and its splitting in the correlation device and crossed Andreev reflection device. **d,** Schematic plot of $dI_{non-local}/dV$ can capture the topological phase transition as an odd function of bias with negative non-local conductance on one side. The black dashed line indicates the topological transition, consistent with the local conductance.

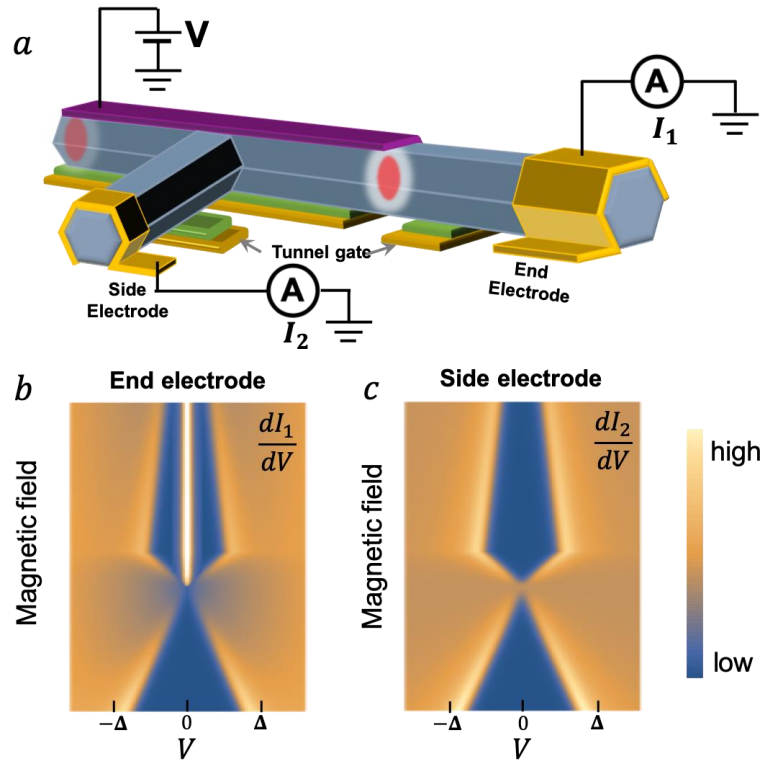

**Figure 4 | Spatially resolving local density of states in a Majorana nanowire**. **a**, Schematic experimental setup for a three terminal T-shape device to resolve the LDOS at the end and middle of a Majorana nanowire. **b**, Schematic plot of $dI_1/dV$ from the end electrode probing the LODS at the wire end. **c**, Schematic plot of $dI_2/dV$ from the side electrode probing the LDOS in the wire middle which shows no zero bias peak.

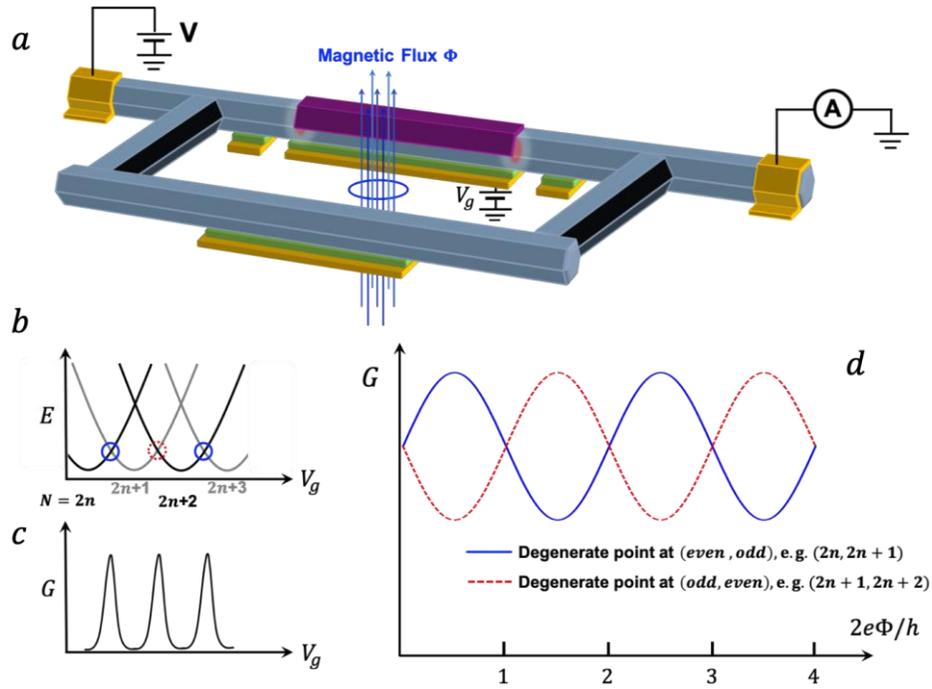

**Figure 5 | Majorana-Fu-Interferometer. a**, Schematic experimental setup of a Majorana-Fu-interferometer device with two coherent electron traveling paths: 1) Majorana island with a finite charging energy in the upper path; 2) semiconductor nanowire in the lower path. **b**, Energy diagram of the island hosting MZMs where N represents the electron number on the island, tuned by $V_g$. **c**, Zero bias conductance of the Majorana island (with the lower path pinched off) shows $1e$-periodic Coulomb blockade peaks, indicating coherent 'teleportation' of single electron processes. **d**, Turning on the lower path as a reference channel, the zero bias conductance shows Ahronov-Bohm oscillations as a function of magnetic flux. The blue and red AB oscillations are for $V_g$ fixed at two successive Coulomb peaks, *i.e.* the two charge degenerate points: even to odd (blue) and odd to even (red). The π phase shift between different parity states can be used for the readout of Majorana qubit.

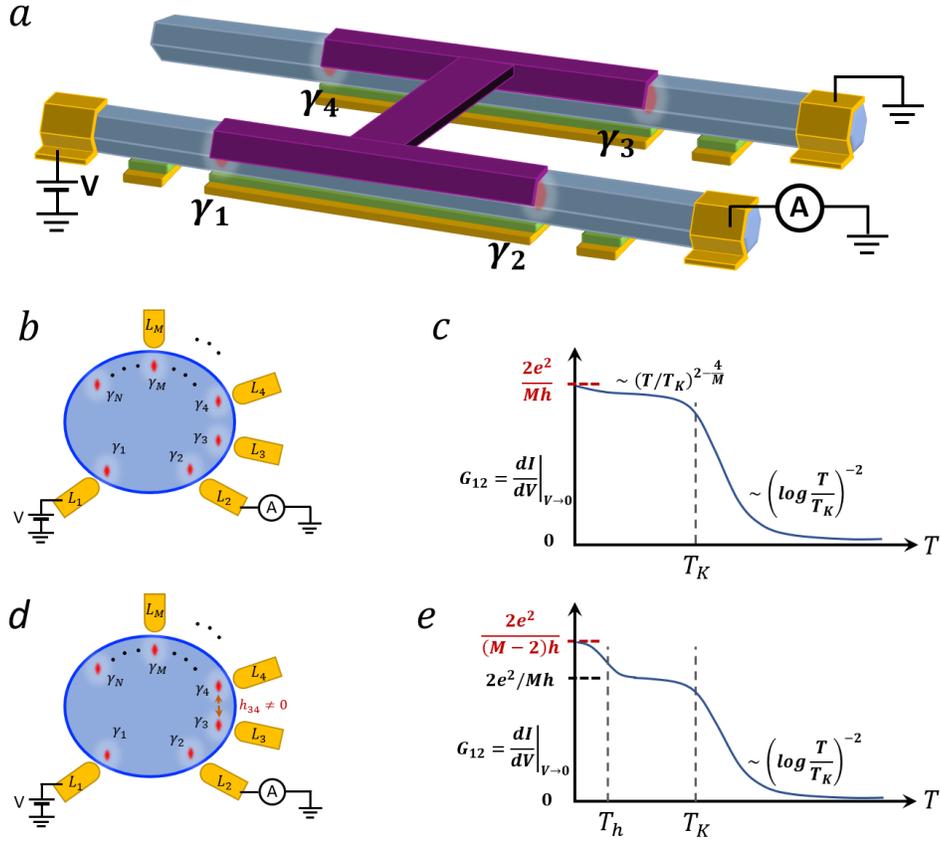

**Figure 6 | Topological Kondo effect. a**, Schematic experimental setup supporting a test of topological Kondo effect with four MZMs and three normal leads. Two parallel semiconductor nanowires are connected by an s-wave superconductor (the "H-shape" junction, purple). **b**, Schematics for a general topological Kondo system with $N$ MZMs and $M$ normal leads, where $N = 4$ and $M = 3$ in **a**. Assuming no spatial overlap between different MZMs, the conductance temperature dependence (**c**) shows a crossover between a weak coupling trivial fixed point at high temperature to a $SO_2(M)$ non-trivial non-Fermi-liquid (NFL) fixed point ($G_{12} = 2e^2/Mh$) at low temperature, separated by the Kondo temperature $T_K$. **d-e**, For the case where two MZMs (e.g. $\gamma_3$ and $\gamma_4$) have a weak hybridization, the system initially tries to crossover to $SO_2(M)$ NFL fixed point, but finally turn to $SO_2(M-2)$ NFL fixed point with $G_{12} = 2e^2/(M-2)h$. The second crossover energy scale is the "Zeeman coupling" $T_h$.

**Author Contributions**: H. Z. conceived the idea. The manuscript was written by H. Z. and D. E. L. with comments from M. W. and L. P. K..

**Author Information**: The authors declare no competing interests.